\documentclass[a4paper,11pt]{article}
\pdfoutput=1 

\usepackage{jheppub} 

\usepackage[T1]{fontenc} 

\usepackage{pdfsync}
\usepackage{mathptmx}
\usepackage[utf8]{inputenc}
\usepackage[T1]{fontenc}
\usepackage{slashed}
\usepackage{times}
\usepackage{amssymb,amsfonts,amsmath,amsthm}
\usepackage{dsfont,bbm}
\usepackage{dcolumn}
\usepackage{epsf}
\usepackage{graphicx}
\usepackage[caption=false]{subfig}
\usepackage{dsfont}
\usepackage{simplewick}
\usepackage{bm}
\usepackage{eucal}
\usepackage{slashed}
\usepackage[active]{srcltx}
\usepackage[usenames]{color}

\title{\boldmath Non-symmetric integral geometry: 
alternative reconstruction procedure with  
dual Radon transform
}

\author[a, b, c]{I.~V.~Anikin}
\affiliation[a]{Bogoliubov Laboratory of Theoretical Physics, JINR, Dubna, Russia}
\affiliation[b]{Institute of Modern Physics, Chinese Academy of Sciences, Lanzhou, P.R. China}
\affiliation[c]{Southern Center for Nuclear Science Theory,
Institute of Modern Physics, Chinese Academy of Sciences, Huizhou, P.R. China}

\emailAdd{anikin@theor.jinr.ru}

\abstract{
In the note, we use the principles of non-symmetric
integral geometry (non-integral geometry), 
related to the  non-symmetric restricted integration measure,
to extend the well-known relations traced from the 
Fourier slice theorem and the dual Radon transform.
The proposed generalization is a very useful for the 
alternative restoration of the outset function.
The latter impacts entirely on the image reconstruction process.
The derived relations confirm the complexity 
of the inversion operator thanks for the symmetry breaking of integration measure. 
}

\begin{document}
\maketitle
\flushbottom

\section{Introduction}
\label{Intro}

Integral geometry has been backed by the symmetry properties of the corresponding
measures on the geometrical spaces. The direct Radon transform (DRT)
as the singular delta-like functional on the well-defined manifold is one
of the main objects in integral geometry. In addition, the integral measure
defined by the differential forms in DRT possesses the certain symmetry
and leads to the invariant measure
\cite{Gelfand:1964, GGV}.

The different integral geometry methods form 
the mathematical tools or basis for computerized tomography (CT), astrophysics and seismology
\cite{Deans}. Indeed, the inverse Radon transforms (IRT)
are being used to reconstruct the image of an object under investigation.
As a rule,
the Radon inversion meets the problems dictated by the certain ill-posedness.
It leads to the necessity to regularize the integral representation of the inverse operator
(the different methods of regularization can be found, for example, 
in  \cite{Anikin:2019oes, Anikin:2024vto}).

Recently, 
the regularization of inverse operator
has been implemented in the form that ensures the universality of inverse Radon transform,
see  \cite{Anikin:2019oes, Anikin:2024vto,Anikin:2025uqv,Anikin:2025lta}.
The universality of inverse Radon transform means the independence of inversion
on the space dimension. 
Notice that the obtained 
universal representation for the Radon inversion has been
derived with the help of the appropriated regularization procedure, used in the generalized function (distribution) theory \cite{Gelfand:1964, GGV, GGG}, and, what is important, 
without referring Courant-Hilbert's identities.

In contrast to 
\cite{Anikin:2019oes, Anikin:2024vto,Anikin:2025uqv,Anikin:2025lta},
the old-fashion Radon inversion
has been realized in the odd and even dimension of space separately 
and independently each other.
The dependence of old-fashion inversion methods on the parity of space dimension can be traced from the fact that 
the angular integration has been always performed over the full region of
variations, {\it i.e.} in the full interval $(0,\, 2\pi)$ 
(see Courant-Hilbert's identities in \cite{Courant-Hilbert}).

As demonstrated, if the original homogeneous outset function, which is under DRT,
is well-localized in the space and does not possess the symmetry properties,
the Radon $\varphi$-angular 
dependence receives the definite restrictions, instead of the full region $(0,\, 2\pi)$.
Due to these angular restrictions, the universal Radon inversion becomes 
a complex transformation. It involves two contributions
$f_S$ and $f_A$, where $f_A$ (this is the additional term compared to the standard inversion)
is induced by the complex integration measure.
This complexity is a very important finding that gives a possibility to extend and to improve the Tikhonov-like regularization
needed for the different practical applications \cite{Anikin:2024vto}.
In addition, the inhomogeneity of the outset function support leads also to the existence of the
additional term $f_A$ independently of the angular restrictions.
We emphasize that all these issues have been initiated by 
the symmetry breaking effects. Meanwhile,
the standard principles of integral geometry have been built up on the 
certain symmetry which is prejudged by the corresponding integration measure 
\cite{Gelfand:1964, GGV, GGG}. 

The presented principles of Radon inversion 
extend and generalize essentially the integral geometry method.
It is worth to notice that
the new principles of the extended integral geometry method are previously 
called as  non-integral geometry \cite{Anikin:2025uqv,Anikin:2025lta}. 
To avoid misunderstanding, we stress that  
the terminologies like ``integral'' and ``non-integral'' are actually corresponding to
 ``complete'' and ``incomplete'' integration measures, respectively. 
 In particular, the use of ``non-integral'' does not mean that the 
 generalized method excludes the integral representations.
For the practical applications, the method of 
the generalized integral geometry 
(non-integral geometry) is more
suitable and adequate because the original objects
corresponding to the outset functions are naturally non-symmetric ones.

As known, the image reconstruction procedure can be implemented  in an alternative way
using not only the Radon inversion. It can be done with the help of the Fourier 
image of dual Radon transforms. 

In the presented paper, 
we use the non-symmetric, extended
integral geometry
to improve the relations emanated from the 
Fourier slice theorem and the dual Radon transform.
The proposed generalization impacts on the image reconstruction 
process \cite{AAS-Mod, Marusina:2006ct}.
It is worth to notice that 
the derived new relations confirm the appearance of complexity 
in the inversion thanks for the symmetry breaking of integration measure. 

\section{Preliminaries}

\subsection{Dual Radon transforms}
\label{Sec-DuRT}

For the convenience, in this section we 
remind the definition of the dual Radon transform given by 
\begin{eqnarray}
\label{Dual-R-1}
\mathcal{R}^\ast[G](\vec{\bf x}) =
\int_{\omega} d\mu(\tau, \varphi) \,G(\tau, \varphi)
\delta\left( \tau - \langle\vec{\bf n}_\varphi,\vec{\bf x}\rangle\right),
\end{eqnarray}
where $G(\tau, \varphi)$ is an arbitrary function,
$d\mu(\tau, \varphi)$ defines the integral measure and
$\vec{\bf n}_\varphi$ implies the unit vector, see for example \cite{Anikin:2024vto}.
 After $\tau$-integration, we obtain that 
\begin{eqnarray}
\label{Dual-R-1-2}
\mathcal{R}^\ast[G](\vec{\bf x}) 
=\int_{\omega_\varphi} (d\varphi) \,G(\langle\vec{\bf n}_\varphi,\vec{\bf x}\rangle, \varphi),
\end{eqnarray}
where, and throughout of this paper, all the relevant numerical normalization factors have 
been incorporated in the integration measure. 
Next, we assume that $G(\tau, \varphi)$ coincides with the Radon image 
$\mathcal{R}(\tau, \varphi)$, {\it i.e.} the dual Radon transform is given by 
$\mathcal{R}^\ast[\mathcal{R}](\vec{\bf x})\equiv \mathcal{R}^\ast\mathcal{R}[f](\vec{\bf x})$. 
As a result, we can write the Fourier image of the dual $\mathcal{R}$-transformation as 
the following:
\begin{eqnarray}
\label{Ft-Rst-1}
{\cal F}[\mathcal{R}^\ast\mathcal{R}[f]](\vec{\bf q})=\int_{-\infty}^{+\infty} (d^2 \,\vec{\bf x}) 
e^{- i \langle \vec{\bf q}, \vec{\bf x}\rangle}
\mathcal{R}^\ast\mathcal{R}[f](\vec{\bf x}),
\end{eqnarray}
and the inverse Fourier transforms are 
\begin{eqnarray}
\label{Inv-Ft-Rst-1}
\mathcal{R}^\ast\mathcal{R}[f](\vec{\bf x})=\int_{-\infty}^{+\infty} (d^2 \,\vec{\bf q}) 
e^{ i \langle \vec{\bf q}, \vec{\bf x}\rangle}
{\cal F}[\mathcal{R}^\ast\mathcal{R}[f]](\vec{\bf q}).
\end{eqnarray}
This representation is very important for our further consideration.

\subsection{Fourier space in the polar system}
\label{Sec-Ft-PolS}

Since a key point of our study is the Fourier slice theorem, we now focus on 
the support domain related to the radial $\lambda$-component 
of the Fourier variables in the polar system.
We begin with the inverse Fourier transform given by 
\begin{eqnarray}
\label{Inv-ft-1}
f(\vec{\bf x})=\int_{-\infty}^{+\infty} (d^n \,\vec{\bf q}) 
e^{i \langle \vec{\bf q}, \vec{\bf x}\rangle}
{\cal F}[f](\vec{\bf q}).
\end{eqnarray}
In the polar system the radial $\lambda$-component of Fourier variables, 
$\vec{\bf q} = \lambda\vec{\bf n}_\varphi $, varies in the standard interval $(0,\, +\infty)$.
In our case, the Fourier radial $\lambda$-component is not restricted because 
the outset function $f(\vec{\bf x})$ with the localized or finite support is only interested,
see App.~\ref{App-1} .
That is, in the polar system we have the following:
\begin{eqnarray}
\label{Inv-ft-2}
f(\vec{\bf x})=\int_{0}^{+\infty} d\lambda\, \lambda  \int_{\omega_\varphi} (d\varphi) 
e^{i \lambda \langle \vec{\bf n}_\varphi, \vec{\bf x}\rangle}
{\cal F}[f](\lambda, \varphi).
\end{eqnarray}
The analytical extension of $\lambda$ up to the full interval $(-\infty,\, +\infty)$ depends 
on the symmetry property of the integration $\varphi$-measure.
Indeed, we can readily see that (\ref{Inv-ft-2}) can be rewritten as 
\begin{eqnarray}
\label{Inv-ft-3}
f(\vec{\bf x})=
\frac{1}{2} \int_{\omega_\varphi} (d\varphi)
\Big\{ 
\int_{0}^{+\infty} d\lambda\, \lambda   
e^{i \lambda \langle \vec{\bf n}_\varphi, \vec{\bf x}\rangle}
{\cal F}[f](\lambda, \varphi) 
- 
\int_{-\infty}^{0} d\tilde\lambda\, \tilde\lambda   
e^{i \tilde\lambda \langle \vec{\bf n}_{\varphi+\pi}, \vec{\bf x}\rangle}
{\cal F}[f](\tilde\lambda, \varphi+\pi)
\Big\}.
\end{eqnarray}
If we suppose that 
\begin{eqnarray}
\label{Full-reg-phi}
\int_{\omega_\varphi} (d\varphi) = \int_{0}^{2\pi} (d\varphi),
\end{eqnarray}
then (\ref{Inv-ft-3}) takes the form of
\begin{eqnarray}
\label{Inv-ft-4}
f(\vec{\bf x})\Big|_{\omega_\varphi-\text{full reg.}}=
\frac{1}{2} \int_{0}^{2\pi} (d\varphi)
\int_{-\infty}^{+\infty} d\lambda\, \big| \lambda \big|   
e^{i \lambda \langle \vec{\bf n}_\varphi, \vec{\bf x}\rangle}
{\cal F}[f](\lambda, \varphi),
\end{eqnarray}
where $\big| \lambda \big|=\lambda \varepsilon(\lambda) \equiv 
\lambda (\theta(\lambda) - \theta(-\lambda))$. However, if we deal with the restricted 
domain of $\varphi$-integration:
\begin{eqnarray}
\label{Res-reg-phi}
\int_{\omega_\varphi} (d\varphi) = \int_{-\pi/2}^{\pi/2} (d\varphi),
\end{eqnarray}
the representation (\ref{Inv-ft-4}) is invalid. Indeed, in the case of (\ref{Res-reg-phi}), we have 
as follows
\begin{eqnarray}
\label{Inv-ft-5}
f(\vec{\bf x})=
\frac{1}{2} 
\Big\{
\int_{-\pi/2}^{\pi/2} (d\varphi)
\int_{0}^{+\infty} d\lambda    
- 
\int_{\pi/2}^{3\pi/2} (d\varphi)
\int_{-\infty}^{0} d\lambda 
\Big\}  \lambda \,
e^{i \lambda \langle \vec{\bf n}_{\varphi}, \vec{\bf x}\rangle}
{\cal F}[f](\lambda, \varphi),
\end{eqnarray}
where the Fourier images in $\Omega_{I+IV}$-region, see the first term of (\ref{Inv-ft-5}), 
and $\Omega_{II+III}$-region, see the second term of (\ref{Inv-ft-5}), are not necessarily 
equal to each other.

Thus, the natural interval of variation for the polar radial $\lambda$-component of 
Fourier variables $\vec{\bf q}$ is given by $(0,\, +\infty)$. The extension of $\lambda$-variable up to the 
full interval is possible provided the special properties related to the integration domain 
of $\varphi$-component which do not match our consideration at all.   

\section{Inverse Fourier slice theorem}
\label{Sec-Inv-FSth}

In this section, we dwell on the inverse Fourier slice theorem which plays the important role 
for the derivation of an alternative inversion procedure to reconstruct the image.
We remind the Fourier slice theorem that states 
 \begin{eqnarray}
\label{FSth-1}
{\cal F}[f](\lambda, \varphi)=
\int_{-\infty}^{+\infty} (d\tau)\,    
e^{ - i \lambda \tau}
\mathcal{R}[f](\tau, \varphi),
\end{eqnarray}
where $\lambda\in (0,\, +\infty)$. Let us now invert the transformation given by 
(\ref{FSth-1}). 
Since in (\ref{FSth-1}) the Fourier radial $\lambda$-component and 
the Radon radial $\tau$-component vary in the different regions ({\it i.e.} in the semi-full and full regions, respectively) we readily obtain that
 \begin{eqnarray}
\label{Inv-FSth-1}
\boxed{
\int_{0}^{+\infty} (d\lambda)\,    
e^{ i \lambda \tilde\tau} {\cal F}[f](\lambda, \varphi)=
\int_{-\infty}^{+\infty} (d\tau)\,    
\frac{i}{\tilde\tau - \tau +i\varepsilon}
\mathcal{R}[f](\tau, \varphi),
}
\end{eqnarray}
where the integral representation of $\delta_+$-function:
 \begin{eqnarray}
\label{Delta-plus}
\delta_+(\tilde\tau - \tau)\equiv \frac{i}{\tilde\tau - \tau +i\varepsilon}=
\int_{0}^{+\infty} (d\lambda)\,    
e^{ i \lambda (\tilde\tau - \tau)},
\end{eqnarray}
 has been used. Hence, one can see that the inverse Fourier slice theorem has more 
 complicated representation compared to the standard inversion.
 As above-mentioned, the standard inversion implies the ideal case where 
 all conjugated variables 
 vary in the full regions owing to the absence of the $\varphi$-angular restrictions. 
 We call the inversion given by (\ref{Inv-FSth-1}) as a generalized inverse Fourier slice theorem. 

\section{Fourier transform of $\mathcal{R}^\ast\mathcal{R}[f]$-function}
\label{Sec-FT-DuR}

We are in a position to consider the Fourier transform of $\mathcal{R}^\ast\mathcal{R}[f]$ in detail. 
We return to the representation (\ref{Ft-Rst-1}) where $\mathcal{R}^\ast\mathcal{R}[f]$-function  
has been replaced by the combination based on the inverse Fourier slice theorem, 
see (\ref{Inv-FSth-1}).
We have 
\begin{eqnarray}
\label{Ft-Rst-w-1}
{\cal F}[\mathcal{R}^\ast\mathcal{R}[f]](\vec{\bf q})=\int_{-\infty}^{+\infty} (d^2 \,\vec{\bf x}) 
e^{- i \langle \vec{\bf q}, \vec{\bf x}\rangle}
\Big\{ 
\int_{\omega} d\mu(\tau, \varphi) \,\mathcal{R}(\tau, \varphi)
\delta\left( \tau - \langle\vec{\bf n}_\varphi,\vec{\bf x}\rangle\right)
\Big\},
\end{eqnarray}
with 
 \begin{eqnarray}
\label{R-FSth-1}
\mathcal{R}(\tau, \varphi)=
\int_{0}^{+\infty} (d\lambda)\,    
e^{ i \lambda \tilde\tau} {\cal F}[f](\lambda, \varphi)
-
\frac{i}{\pi}
\int_{-\infty}^{+\infty} (d\eta)\,    
\frac{\mathcal{P}}{\tau - \eta}
\mathcal{R}[f](\eta, \varphi).
\end{eqnarray}
The representation (\ref{R-FSth-1}) can be immediately derived from (\ref{Inv-FSth-1}) 
provided the real and imaginary parts are separetly singled out.
After simple algebra, we finally derive that 
\begin{eqnarray}
\label{Ft-Rst-w-2}
\boxed{
{\cal F}[\mathcal{R}^\ast\mathcal{R}[f]](\vec{\bf q})=
\frac{1}{\big|  \vec{\bf q} \big|} {\cal F}[f](\vec{\bf q}) -
\frac{i}{\pi}
\big[ \mathcal{R}[f] \circledast Q  \big](\vec{\bf q})
},
\end{eqnarray}
where, by definition,
\begin{eqnarray}
\label{Ft-Rst-w-3}
\big[ \mathcal{R}[f] \circledast Q  \big](\vec{\bf q})=
\int_{\omega_\varphi} (d\varphi)
\int_{-\infty}^{+\infty} (d\eta) Q(\eta, \varphi; \vec{\bf q})\,
\mathcal{R}[f](\eta, \varphi)
\end{eqnarray}
with 
\begin{eqnarray}
\label{Ft-Rst-w-4}
Q(\eta, \varphi; \vec{\bf q})=
\int_{-\infty}^{+\infty} (d^2 \vec{\bf x}) 
e^{ - i \langle \vec{\bf q}, \vec{\bf x}\rangle} 
\frac{\mathcal{P}}{\langle\vec{\bf n}_\varphi,\vec{\bf x}\rangle-\eta}.
\end{eqnarray}
Owing to its second term, the relation (\ref{Ft-Rst-w-2}) is completely new one 
and it is presented at the first time.
Therefore, we derive the generalized representation for the Fourier transform of the dual Radon image,
see (\ref{Ft-Rst-w-2}). This generalized representation plays an important role for the
derivation of an alternative inversion procedure used in the image reconstruction process.

\section{Alternative and extended reconstruction process}
\label{Sec-Al-RP}

The new and extended representation (\ref{Ft-Rst-w-2}), presented in the preceding section,
 can be used for 
the derivation of an alternative inversion procedure for the image reconstruction.
Indeed, we first focus on the standard inverse Fourier transform 
given by (\ref{Inv-ft-1}) in $\mathbb{R}^2$. Then, we insert the representation for the Fourier image of 
$f$-function, extracted from (\ref{Ft-Rst-w-2}), into (\ref{Inv-ft-1}). We finally get the following
\begin{eqnarray}
\label{Inv-ft-ww-1}
\boxed{
f(\vec{\bf x})=\int_{-\infty}^{+\infty} (d^2 \,\vec{\bf q}) 
e^{i \langle \vec{\bf q}, \vec{\bf x}\rangle}
\big|  \vec{\bf q} \big|
\Big\{
{\cal F}[\mathcal{R}^\ast\mathcal{R}[f]](\vec{\bf q})
+
\frac{i}{\pi}
\big[ \mathcal{R}[f] \circledast Q  \big](\vec{\bf q})
\Big\}.
}
\end{eqnarray}
Notice that, in this representation, the complexity is being manifested as well as in   
\cite{Anikin:2019oes, Anikin:2024vto,Anikin:2025uqv,Anikin:2025lta}.
In addition, the complexity can be also extracted from $Q$-function. Indeed, we have 
the following
\begin{eqnarray}
\label{Ft-Rst-w-5}
Q(\eta, \varphi; \vec{\bf q})\Big|_{\vec{\bf q}=\lambda \vec{\bf n}_\varphi}=
\int_{\omega_\phi} (d\phi) \int_{0}^{+\infty} (dr) \, r\, 
e^{ - i \lambda r \cos(\varphi - \phi)} 
\frac{\mathcal{P}}{r \cos(\varphi - \phi) -\eta},
\end{eqnarray}
where the polar system for the position space, 
$\vec{\bf x}=r\, \vec{\bf n}_\phi=(r,\, \phi)$, has been used.
Hence, we can readily see that 
\begin{eqnarray}
\label{Ft-Rst-w-5}
Q(\eta, \varphi; \vec{\bf q})\Big|_{\vec{\bf q}=\lambda \vec{\bf n}_\varphi}\sim
\int_{0}^{+\infty} (dr) \,  
e^{ - i C r } \frac{r}{r - B} \Rightarrow
\int_{0}^{+\infty} (dr) \,  
e^{ - i C r }= \delta_{-}(C) \in \mathbb{C}.
\end{eqnarray}
We thus derive the new and extended representation for the alternative inversion procedure 
needed for the image reconstruction.

\section{Conclusions}
\label{Cons}

In conclusion, we have demonstrated that the corresponding $\varphi$-angular
restrictions (for example, $\varphi\in (-\pi/2,\, \pi/2)$ as in \cite{Anikin:2019oes, Anikin:2024vto,Anikin:2025uqv,Anikin:2025lta}) 
lead to the important impacts 
on the several inferences which have never been studied previously. 
Namely, the derived inverse Fourier slice theorem becomes to be improved in comparison with  
 the standard inversion where there is no the restriction on the angular component 
 and all the conjugated variables 
 vary in the full regions. In other words, 
 we have first presented the  generalized inverse Fourier slice theorem, see (\ref{Inv-FSth-1}).
Then, we have derived the generalized representation for the Fourier transform of the dual Radon image (see (\ref{Ft-Rst-w-2})) which is a very important one to obtain
the alternative inversion procedure. 
Also, the new and extended representation for the alternative inversion procedure 
needed for the image reconstruction has been received, see (\ref{Inv-ft-ww-1}).

\section*{Acknowledgements}
We thank A.I.~Anikina, D.~Karlovets, V.A.~Osipov, I.~Pavlov, A.~Shchepkin,  O.I.~Streltsova, N.A.~Tu\-rin and D.~Zhu for useful and illuminating discussions.
Also, the author thanks his colleagues  from
the Southern Center for Nuclear Science Theory, Huizhou, China
for a very warm hospitality. Our special thanks go to X.~Chen for the creation of 
excellent group atmosphere.   
The work has been supported in part by PIFI 2026PVA0068
Program.

\section*{Conflict of interest}

The author declares no conflict of interest.

\section*{Data Availability Statement}
This manuscript has no associated data or the data will not be deposited.

\appendix
\renewcommand{\theequation}{\Alph{section}.\arabic{equation}}
\section*{Appendix}


\section{Fourier transforms of the well-localized functions}
\label{App-1}

Let us consider the Fourier transform of the function $f(\vec{\bf x})$:
\begin{eqnarray}
\label{ft-1}
{\cal F}[f](\vec{\bf q})=\int_{-\infty}^{+\infty} (d^n \,\vec{\bf x}) 
e^{-i \langle \vec{\bf q}, \vec{\bf x}\rangle}
f(\vec{\bf x}).
\end{eqnarray}
For the simplicity of reasoning but without losing of generality, 
the function $f(\vec{\bf x})$ is assumed to be $f(x, 0)\equiv f(x)=e^{-x^2}$ with $x\in \mathbb{R}^1$.
In other words, we study the Fourier transformation for the well-localized one-dimensional function $f(x)$.
Then its Fourier image reads
\begin{eqnarray}
\label{dir-ft-1}
{\cal F}[f](q) = \int_{-\infty}^{+\infty} (dx) e^{-(x+iq/2)^2  - q^2/4}.
\end{eqnarray}
To calculate this integral it is useful to make the following replacement:
$z=x+iq/2$. Hence, for the integration limits, we have two cases:
\begin{itemize}
\item ${\it (a)}$ a new integration variable $z$ is a real one due to 
$q< M < \infty$. That is,  
$q$ is rather a small (or finite) Fourier variable
and it can be neglected.
Moreover, this $z$ varies into the interval $(-\infty, \, \infty)$;
\item ${\it (b)}$ a new integration variable $z$ becomes a complex one and varies also in the infinite interval 
provided the Fourier variable $q\sim[\infty]$, {\it i.e.} $q\sim M \to \infty$ ($q$ is a large enough).   
\end{itemize}

By direct calculations of (\ref{dir-ft-1}), one can readily see that the case ${\it (a)}$ leads to the 
well-localized Fourier image:
\begin{eqnarray}
\label{dir-ft-2}
{\cal F}[f](q) = N e^{-q^2/4}.
\end{eqnarray} 
While the case ${\it (b)}$ corresponds to the Fourier image of the well-localized function that is 
not localized at all.
Indeed, we have the following:
\begin{eqnarray}
\label{dir-ft-3}
{\cal F}[f](q) = \int_{-\infty}^{+\infty} (dx) e^{-x^2}  \cos(qx)
\end{eqnarray} 
and 
\begin{eqnarray}
\label{dir-ft-4}
\lim_{q\to\infty}{\cal F}[f](q) \sim 
\lim_{M\to\infty}\cos(M) \not= 0.
\end{eqnarray} 
The other similar example is given by  the characteristic function 
in $\mathbb{R}^1$:
$f(x)=\Theta\big( |x | \le a \big)$.
Having calculated, in this case the Fourier image reads
\begin{eqnarray}
\label{dir-ft-5}
{\cal F}[f](q) = \int_{-a}^{+a} (dx) \cos(qx)
\sim 
\lim_{M\to\infty}\frac{\sin(M)}{M} \not= 0.
\end{eqnarray} 
One can see that (\ref{dir-ft-5}) decreases fast but it is not equal to zero.
This fact becomes to be very important while we deal with the inclusion of the noise effect. 

As a result, the well-localized functions, $e^{-x^2}$ or $\Theta\big( |x | \le a \big)$, 
can give conditionally non-localized its Fourier images.
At the same time, the origin function $e^{-x^2}$ can lead to the 
localized Fourier image owing to the restricted (or finite) Fourier variable $q$. 


\end{document}